\documentclass[twocolumn,showpacs,floatfix,superscriptaddress]{revtex4}
\usepackage{latexsym}
\usepackage{graphics}
\usepackage{epsf}
\usepackage{epsfig}
\usepackage{graphicx}
\usepackage{dcolumn}
\usepackage{bm}
\usepackage{latexsym}
\usepackage{booktabs}
\usepackage{amssymb}
\usepackage[english]{babel}
\usepackage{hyperref}
\usepackage{amsmath}

\begin{document}
\title{Universality class of triad dynamics on a triangular lattice}

\author{Filippo Radicchi}\email[]{f.radicchi@iu-bremen.de}
\affiliation{School of Engineering and Science , International
University Bremen , P.O.Box 750561 , D-28725 Bremen , Germany.}
\author{Daniele Vilone}\email[]{d.vilone@iu-bremen.de}
\affiliation{School of Engineering and Science , International
University Bremen , P.O.Box 750561 , D-28725 Bremen , Germany.}
\author{Hildegard Meyer-Ortmanns}\email[]{h.ortmanns@iu-bremen.de}
\affiliation{School of Engineering and Science , International
University Bremen , P.O.Box 750561 , D-28725 Bremen , Germany.}

\noindent

\begin{abstract}
We consider triad dynamics as it was recently considered by Antal
\emph{et al.} [T. Antal, P. L. Krapivsky, and S. Redner, Phys.
Rev. E {\bf 72} , 036121 (2005)] as an approach to social balance.
Here we generalize the topology from all-to-all to a regular one
of a two-dimensional triangular lattice. The driving force in this
dynamics is the reduction of frustrated triads in order to reach a
balanced state. The dynamics is parameterized by a so-called
propensity parameter $p$ that determines the tendency of negative
links to become positive. As a function of $p$ we find a phase
transition between different kind of absorbing states. The phases
differ by the existence of an infinitely connected (percolated)
cluster of negative links that forms whenever $p\leq p_c$.
Moreover, for $p\leq p_c$, the time to reach the absorbing state
grows powerlike with the system size $L$, while it increases
logarithmically with $L$ for $p > p_c$. From a finite-size scaling
analysis we numerically determine the critical exponents $\beta$
and $\nu$ together with $\gamma$, $\tau$, $\sigma$. The exponents
satisfy the hyperscaling relations. We also determine the fractal
dimension $d_f$ that fulfills a hyperscaling relation  as well.
The transition of triad dynamics between different absorbing
states belongs to a universality class with new critical
exponents. We generalize the triad dynamics to four-cycle dynamics
on a square lattice. In this case, again there is a transition
between different absorbing states, going along with the formation
of an infinite cluster of negative links, but the usual scaling
and hyperscaling relations are violated.
\end{abstract}


\maketitle

\section{Introduction}
Recently Antal \emph{et al.} \cite{antal} proposed a so-called triad
dynamics to model the approach of social balance \cite{heider,cartwright,roberts}. An essential
ingredient in the algorithm is the reduction of frustration in the
following sense. We assign a value of $+1$ or $-1$ to a link (or bond) in the all-to-all topology
if it connects two individuals who are friends or enemies, respectively. We call the sign $\pm 1$
of a link its spin. If the product of links along the boundary of
a triad is negative, the triad is called frustrated (or
imbalanced), otherwise it is called balanced (or unfrustrated).
The state of the network is called balanced if all triads are
balanced. The algorithm depends on a parameter $p \in [0,1]$,
called propensity parameter. It determines the tendency of the system to
reduce frustration via flipping a negative link to a positive one.
For an all-to-all topology Antal \emph{et al.} predict a
transition from an imbalanced non-absorbing stationary state for
$p<1/2$ to a balanced absorbing state for $p\geq 1/2$. Here the
dynamics is motivated by social applications so that the notion of
frustration from physics goes along with frustration in the
psychological sense. The mathematical criterion for checking the status of frustration is the same.\\
In a recent paper \cite{radicchi} we generalized the triad dynamics
in two aspects. The first generalization refers to a k-cycle
dynamics which contains the triad dynamics for $k=3$. Here it
turned out that the main difference comes from the difference of
whether $k$ is even or odd, since the phase structure is symmetric
about $p=1/2$ for $k$ even. Even in the infinite-volume limit
there are only absorbing states, apart from the transition point at
$p=1/2$. The second generalization concerned the network topology
from all-to-all connections to a diluted network. We studied the
phase structure as a function of the propensity $p$ and the
dilution. As it turned out, the diluted k-cycle dynamics can be
mapped on a certain satisfiability problem in computer science,
the so-called $k$-XOR-SAT problem \cite{garey}, and socially balanced states in
one problem correspond to all logical constraints satisfied in the
$k$-XOR-SAT problem. In both models we have phases of imbalanced
non-absorbing states in the infinite-volume limit, separated by a
phase transition from phases of balanced or absorbing states. In a
finite volume one only observes balanced states, but as a remnant
of the infinite-volume phase structure, the time to reach the
absorbing states differs in a characteristic way.\\
In this paper we study triad dynamics on a two-dimensional
triangular lattice and $4$-cycle dynamics (called tetrad dynamics)
on a square lattice. {}From the interpretation as an approach to
social balance, triad or tetrad dynamics on a regular topology are
not more realistic than on an all-to-all topology. Still, triad
dynamics shows interesting features in terms of a percolation
transition if we compare snapshots of frozen states for different
values of $p$. Also for tetrad dynamics we observe a transition
between different absorbing states, but the description in terms
of a percolation transition fails. As we shall see, due to the
restrictive topology, imbalanced triads and tetrads allow only two
elementary processes: either they diffuse or they annihilate each
other. As a result, the system always approaches a balanced
absorbing state; nevertheless we observe a transition as a
function of the propensity parameter $p$, this time between
different absorbing states. The difference is characterized by the
presence or absence of an infinite cluster of connected unfriendly
(negative) links and for triad dynamics by the time to reach these
frozen states. The parameter $p$ should not be confused with the
occupation probability of a single bond with a positive sign as it
usually used in connection with bond percolation. Therefore, here
for $p\leq p_c$ we observe an infinite cluster of negative links,
while such a cluster is absent for $p > p_c$. For triad dynamics
the time to reach the frozen state grows logarithmically with the
system size for $p > p_c$ and as a power of the size for $p\leq
p_c$. In contrast, for tetrad dynamics it grows as a power of the
system size only at the transition point, while it grows
logarithmically above and below $p_c$. For triad dynamics we
numerically determine the critical exponents from a finite-size
scaling analysis of the frozen patterns. We analyze the
probability $P_\infty$ for a link to belong to the infinite
cluster that serves as an order parameter as well as the order
parameter susceptibility and the distribution of clusters of
finite size $s$. The critical exponents $\beta$, $\nu$, $\gamma$,
$\sigma$, $\tau$ fulfill the usual hyperscaling relations as well
as the fractal dimension $d_f$. The critical exponents turn out to
be new and completely different from those of standard percolation
in two dimensions \cite{staufer,havlin}.
\\
The outline of the paper is the following. In section \ref{sec2}
we define the triad dynamics on a triangular lattice. In section
\ref{sec3} we analyze the finite-volume dependence of the frozen
states as a function of the propensity parameter $p$ and present
the results for the critical exponents as well as for the fractal
dimension. Section \ref{sec4} deals with tetrad dynamics on a
square lattice for which the hyperscaling relations are violated. In section
\ref{sec5} we summarize the conclusions. \noindent

\section{Triad dynamics}\label{sec2}
\begin{figure}
\includegraphics*[width=0.47\textwidth]{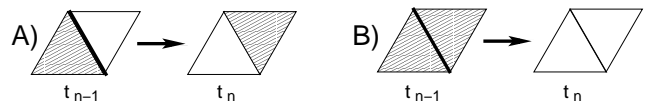}
\caption{Local dynamics of one update event. The imbalanced triads
are represented as filled triangles, while the balanced ones as
empty triangles. The shared link is represented as bold, this is
the link involved in the update event. A) When the update event
flips the link shared by one imbalanced and one balanced triad we
have diffusion. B) When the update event flips the link shared by
two imbalanced triads we have annihilation.} \label{fig:expl1}
\end{figure}

\begin{figure}
\Large{A)}
\\
\includegraphics*[width=0.42\textwidth]{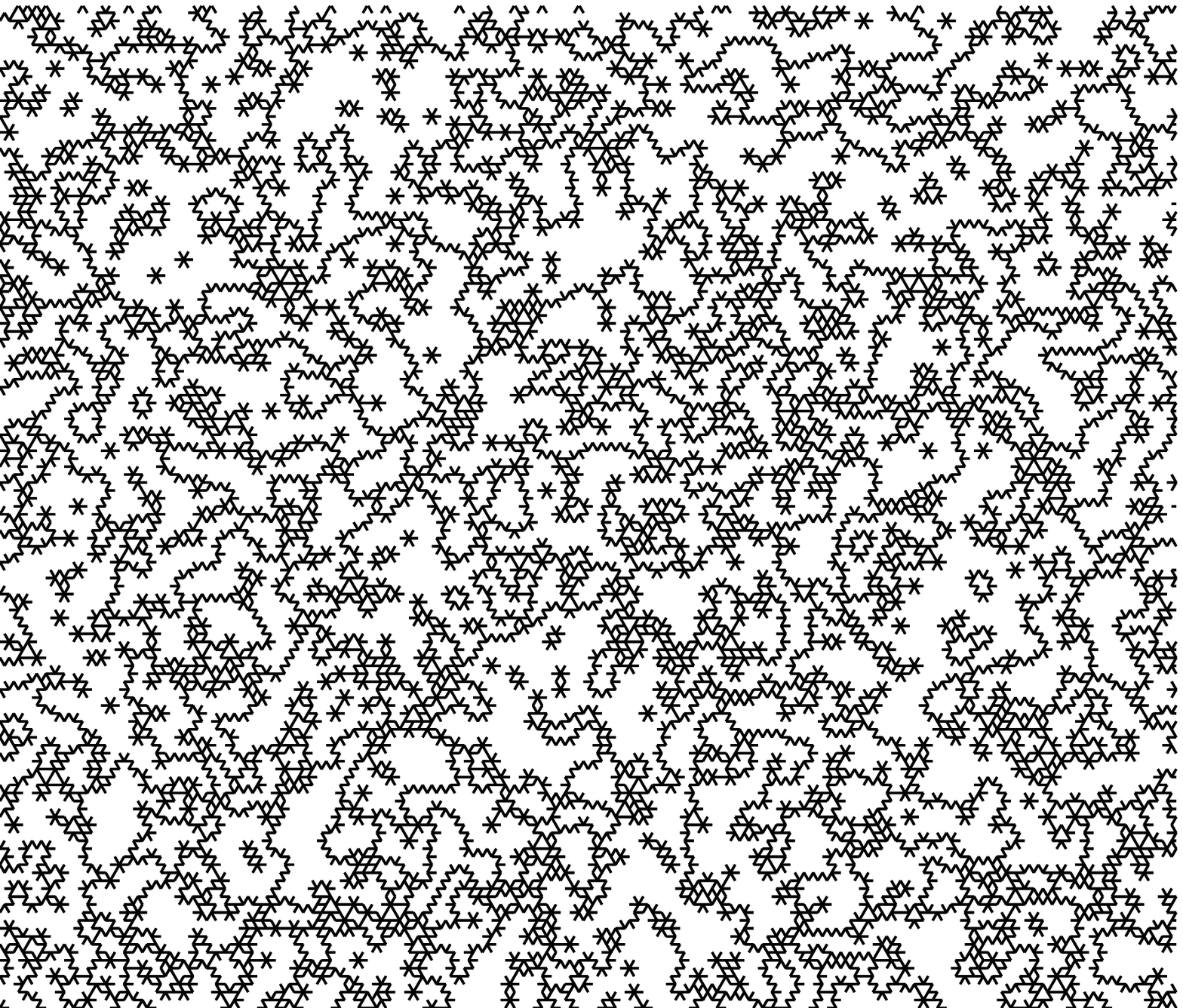}
\\
\vspace{0.3cm}
\Large{B)}
\\
\includegraphics*[width=0.42\textwidth]{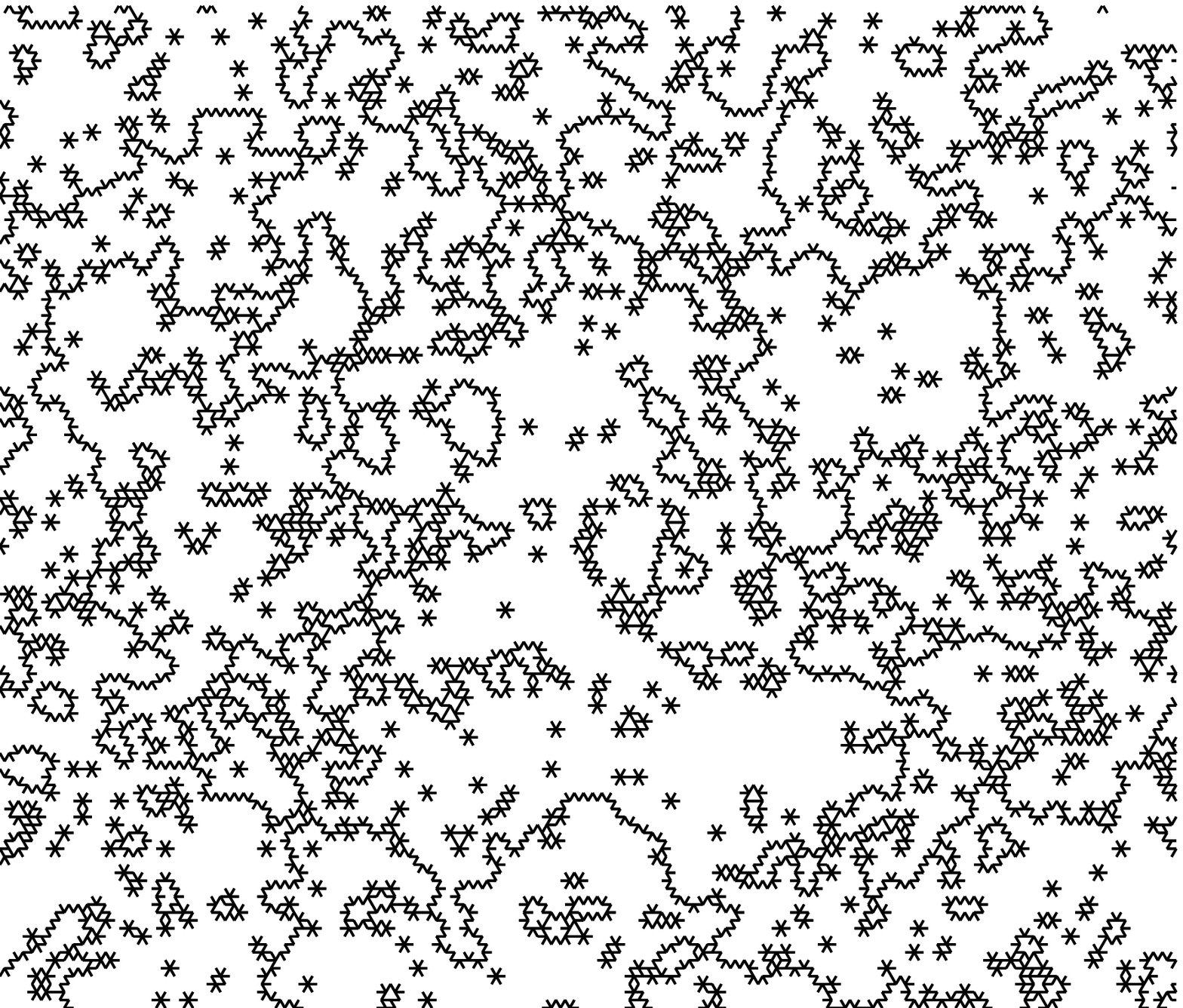}
\\
\vspace{0.3cm}
\Large{C)}
\\
\includegraphics*[width=0.42\textwidth]{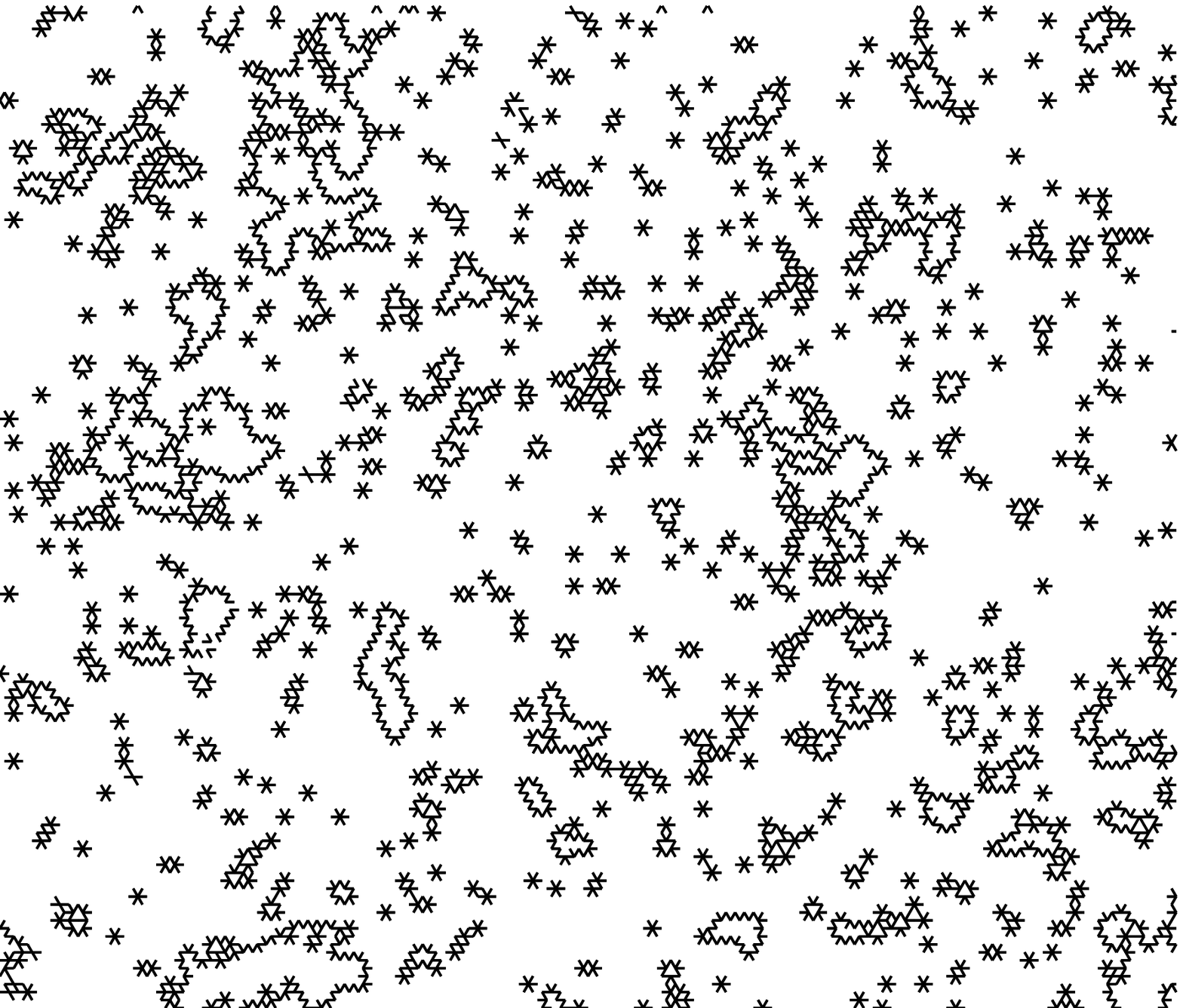}
\caption{Typical frozen configurations for a triangular lattice
with periodic boundary conditions, with linear size $L=129$ and
for different values of $p$: A) $p=0.44$ , B) $p=p_c=0.4625$ , C)
$p=0.48$. In the plots only negative links are shown.}
\label{fig:expl}
\end{figure}

Our dynamical system is defined on an undirected graph (network)
composed of nodes and links. Each link $(i,j)$ between the nodes
$i$ and $j$ takes spin values $\sigma_{(i,j)}=-1$ or
$\sigma_{(i,j)}=+1$ if the nodes $i$ and $j$ are  ``enemies'' or
``friends'', respectively. A triad [$\triangle$]  $(i,j,k)$ is
characterized by the values assigned to its three links $(i,j)$,
$(j,k)$ and $(k,i)$. We have four types of triads,
depending on the number of negative links they contain in their
boundary: $\triangle_0, \triangle_1, \triangle_2$ and
$\triangle_3$ , where the subscript stands for the number of
negative (or ``unfriendly'') links. We use the standard notion of
\emph{social balance} as proposed in \cite{heider,cartwright} and
apply this notion to triads. The sign of a triad is defined as the
product of the spins assigned to the links of the triad. A triad
is considered as ``balanced'' or ``unfrustrated'' if its sign is
positive, otherwise it is called ``imbalanced'' or ``frustrated''.
The triads $\triangle_0$ (all friends) and $\triangle_2$ (two
friends have the same enemy) are balanced, while the triads
$\triangle_1$ and $\triangle_3$ are imbalanced. The network itself
is called balanced if and only if all triads belonging to the
network are balanced.
\\
As in \cite{antal,radicchi} we perform a local unconstrained dynamics in
order to reduce the frustration of the network. As it turns out,
the local algorithm always drives the network to a fully balanced
state without frustrated triads, but the time it needs for
reaching the frozen state depends on the choice of parameters.
At each update event one triad is selected at random.
If the selected triad is balanced (type $\triangle_0$ or $\triangle_2$) nothing happens.
If the selected triad is imbalanced (type $\triangle_1$ or $\triangle_3$)
it is updated into a balanced one by flipping one of its link.
In particular a triad $\triangle_1$ is changed with probability $p$
into a triad $\triangle_0$ (flipping the only negative link of the
triad $\triangle_1$), and it is changed with probability $1-p$
into a triad  $\triangle_2$ (choosing at random one of the two
positive links belonging to the triad $\triangle_1$ and inverting
the spin to a negative value). A triad $\triangle_3$ is changed
into a triad $\triangle_2$ with probability $1$,  choosing at
random one of its three negative links and reversing its spin to a
positive sign. We summarize the updating rules of the local
algorithm in the following scheme:
\begin{equation}
\triangle_0 \;\; \xleftarrow[p]{} \;\;
\triangle_1 \;\; \xrightarrow[1-p]{} \;\;
\triangle_2 \;\; \xleftarrow[1]{} \;\; \triangle_3 \;\;\;.
\label{eq:triad_rules}
\end{equation}
One time unit has passed when the number of single update events equals
the total number of links $M$ of the network.
\\
We study the triad dynamics on two-dimensional triangular lattices
with periodic boundary conditions.
We characterize a triangular lattice using its linear size $L$.
The total number of sites in the lattice is $N=L(L-1)$. The
term $L-1$ results from the periodic boundary conditions. The total
number of links of the lattice is $M=3N$, while the total number
of triads is $N_{\triangle}=2N$. In particular each link is shared only by
two nearest-neighbor triads, so that the triad dynamics cannot increase
the total number of imbalanced triads: a single update changes one
selected imbalanced triad into a balanced one, while it modifies
the other triad, sharing the same updated link, either from balanced to
imbalanced [Figure \ref{fig:expl1}A)] or from
imbalanced to balanced [Figure
\ref{fig:expl1}B)]. The former we call
diffusion, because the imbalanced triad diffuses, the
latter annihilation between two imbalanced triads.
Obviously, for a finite-size system we always observe a frozen
configuration as the stationary state (see Figure \ref{fig:expl}),
independently on the initial configuration. {}From now on we focus
on these frozen configurations and study their geometrical
properties by using the standard tools of percolation theory
\cite{staufer}. It should be noticed, however, that the
dynamical parameter $p$ of the triad dynamics is very different
from the occupation probability as it is
defined in percolation theory, where it is also called $p$.

\section{Numerical simulations to determine the critical threshold and the critical exponents}
\label{sec3} In order to study the geometry of the frozen
configurations, we consider the distribution of negative links on
the lattice. In particular we numerically compute the probability
that a negative link belongs to an infinite cluster $P_\infty =
M^-_\infty / M$, as the ratio of the number of links belonging to
the largest cluster of connected negative links $ M^-_\infty$ to
the total number of links $M$.

\subsection*{Critical propensity parameter $p_c$}
\begin{figure}
\includegraphics*[width=0.47\textwidth]{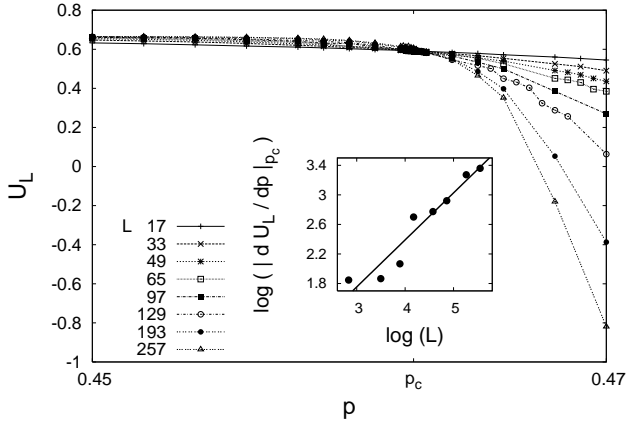}
\caption{Binder cumulant $U_L$ as a function of the propensity
parameter $p$. The main plot shows $U_L$ in the vicinity of the
critical point. The numerical results for different linear lattice
sizes $L$ have a common intersection at $p_c \simeq 0.463$. The
inset shows the $log$-$log$ plot of $\left| dU_L/dp
\right|_{p_c}$. {}From the relation $\left| dU_L/dp \right|_{p_c}
\sim L^{1/\nu}$ we find $1/\nu=0.64(7)$ (solid line).}
\label{fig:cum}
\end{figure}

Let us first determine the critical point $p_c$. Binder's cumulant
\cite{binder}, defined as the $4$-th order reduced cumulant of the
probability distribution
\begin{equation}
U_L= 1- \frac{\langle P_\infty^4 \rangle}{3\langle P_\infty^2
\rangle^2}\;, \label{eq:cumulant}
\end{equation}
allows us to determine the critical point $p_c$ without tuning any
parameter. In Eq.(\ref{eq:cumulant}) $\langle \cdot \rangle$
stands for the average over all realizations of the distribution.
It is known that the Binder cumulant  should satisfy the scaling
relation \cite{binder}
\begin{equation}
U_L = \tilde{U} \left[\left(p-p_c \right)L^{1/\nu} \right] \;
, \label{eq:cumulant2}
\end{equation}
with $\tilde{U}(\cdot )$ a universal function. {}From Figure
\ref{fig:cum} it is obvious that the Binder cumulants for
different linear sizes of the lattice $L$ have a common
intersection at $p=0.4625(5)$, so that we use
$p_c=0.4625$ as the critical propensity parameter in the
following analysis. The numerical results of Figure
\ref{fig:cum} are extracted from numerical simulations for $L =
17$, $33$, $49$, $65$, $97$, $129$, $193$ and $257$. The averages
are taken over several frozen configurations as they are reached
as absorbing states of the triad dynamics, starting from initial
conditions where each link is randomly assigned a value of $+1$ or
$-1$ with the same probability $1/2$. The number of
realizations is $10^4$ for values of $L$ up to $65$ and $10^3$ for
larger values of $L$.

\subsection*{Critical exponent $\nu$}
In order to extract the value of the critical exponent $\nu$ that
characterizes the divergence of the correlation length $\xi$ in
the vicinity of the critical point, we consider the absolute value
of the first derivative of the Binder cumulant calculated at
$p_c$, $\left| dU_L/dp \right|_{p_c}$. We expect from
Eq.(\ref{eq:cumulant2}) to have $\left| dU_L/dp \right|_{p_c} \sim
L^{1/\nu}$. This relation is actually satisfied for $1/\nu =
0.64(7)$ as it is seen in the inset of Figure \ref{fig:cum}, from
which $\nu=1.6(2)$. For the numerical estimate of the derivative
of $U_L$ we use $\left| dU_L/dp \right|_{p_c} \; = \; \left|
\sum_{q=1}^Q q \; a_{L,q} \; p_c^{q-1} \right|$ where the
coefficients $a_{L,q}$ are obtained by interpolating the cumulant
$U_L$ with a polynomial $U_L \; = \; \sum_{q=0}^Q \; a_{L,q} \;
p^q$. Here we choose $Q=5$ and restrict the interpolation to the
interval $[0.46,0.465]$.

\subsection*{Critical exponent $\beta$}
\begin{figure}
\includegraphics*[width=0.47\textwidth]{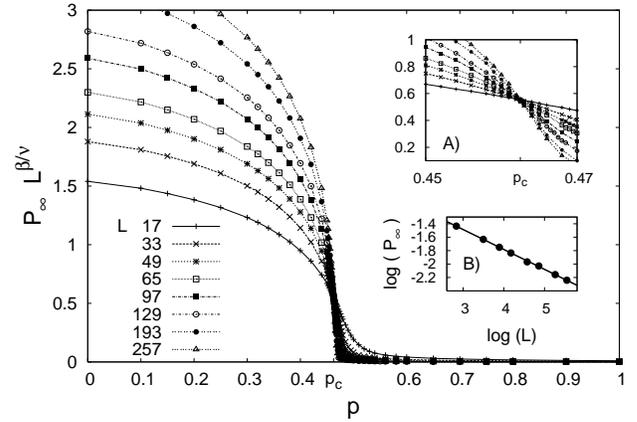}
\caption{Probability $P_\infty$ that a bond belongs to the largest
cluster as a function of $p$ and for different linear sizes of the
lattice. We perform a finite-size scaling for determining the
critical point $p_c=0.4625(5)$ [see the zoom around the critical
point in the inset A)] with the critical exponents  $\beta/\nu =
0.297(3)$ (solid line)  as  is shown  in the inset B), where
$P_\infty$ at the critical point is plotted as a function of the
linear size of the lattice $L$. The numerical simulations  are the
same as in Figure \ref{fig:cum}.} \label{fig1}
\end{figure}

\begin{figure}
\includegraphics*[width=0.47\textwidth]{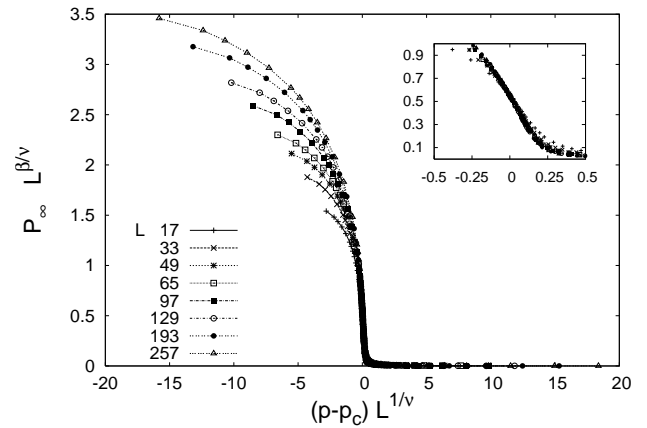}
\caption{Same data as in Figure  \ref{fig1}, but now the abscissa
is rescaled according to $(p-p_c)L^{1/\nu}$ with $1/\nu=0.64$. The
inset shows a zoom around zero. The numerical simulations are the
same as in Figure \ref{fig:cum}.} \label{fig2}
\end{figure}

According to percolation theory, the probability $ P_\infty $
satisfies the following finite-size scaling relation
\begin{equation}
P_{\infty}  = L^{-\beta/\nu} \; \tilde{P} \; \left[ \left(p-p_c
\right) L^{1/\nu} \right]\;, \label{eq:scaling}
\end{equation}
where $p_c$ is the critical value of the propensity parameter $p$
for which the phase transition occurs. (In the infinite-volume
limit we have $P_\infty =1$ for $p \leq p_c$, while $P_\infty =0$ for
$p>p_c$.) $\tilde{P}(\cdot )$ is a universal function. A
plot of $P_\infty L^{\beta/\nu}$ as a function of $p$ and for
different values of $L$ is shown in Figure \ref{fig1}. For a value
of $\beta/\nu=0.297(3)$ [see inset B)] all curves have an
intersection in $p_c=0.4625(5)$ as it is more obvious from the
inset A). Using for $\nu$ the value calculated so far, we obtain $\beta=0.46(5)$.
\\
If we rescale the abscissa  as $(p-p_c)L^{1/\nu}$ with
$p_c=0.4625$ and $1/\nu=0.64$, all curves for different values of
$L$ collapse into one (see Figure \ref{fig2}). This is true
especially close to zero as the inset of Figure \ref{fig2} clearly
shows.

\subsection*{Critical exponent $\gamma$}
\begin{figure}
\includegraphics*[width=0.47\textwidth]{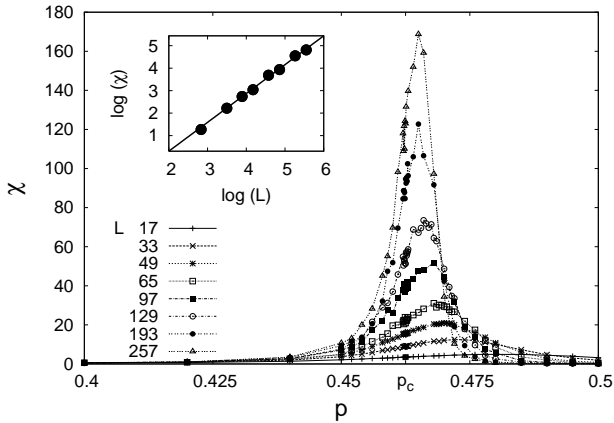}
\caption{Susceptibility $\chi$ of $P_\infty$ as a function of $p$.
In the main plot
we zoom the region around the critical point $p_c$, while in the
inset we plot $\chi$ at the critical point as a function of $L$
leading to $\gamma/\nu=1.28(3)$ (solid line). The numerical
simulations are the same as in Figure \ref{fig:cum}.}
\label{fig2aa}
\end{figure}

\begin{figure}
\includegraphics*[width=0.47\textwidth]{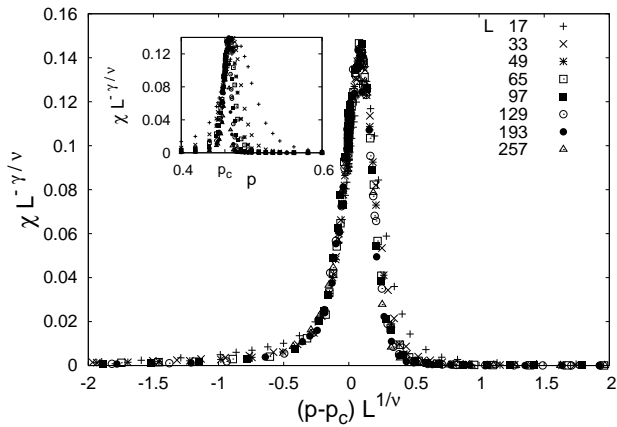}
\caption{finite-size scaling for the susceptibility $\chi$. The
data are the same as in Figure  \ref{fig2aa}. In the inset we zoom
the function $\chi L^{-\gamma / \nu}$ around the critical point
$p_c$ , while in the main plot we rescale the abscissa as $\left(p
-p_c \right) L^{1/\nu}$. The critical exponents are chosen as $\gamma/\nu=1.28$ and $1/\nu=0.64$.} \label{fig2a}
\end{figure}

Moreover, in Figure  \ref{fig2aa} we plot the susceptibility
\begin{equation}
\chi = M \left[ \langle P_\infty^2 \rangle -  \langle P_\infty
\rangle^2 \right]\;\;\; ,\label{eq:susc}
\end{equation}
in which we only show a zoom around $p=p_c$, while in the inset we
plot the value of $\chi$, at the critical point, as a function of
the linear size of the lattice $L$. From the inset we find
$\gamma/\nu=1.28(3)$, because $\chi \sim L^{\gamma/\nu}$ at $p_c$,
therefore $\gamma=2.0(3)$.
The susceptibility should satisfy the finite-size scaling relation
\begin{equation}
\chi = L^{\gamma / \nu} \; \tilde{\chi} \left[\left(p -p_c \right) L^{1/\nu} \right] \;\;\; ,
\label{eq:scaling_susc}
\end{equation}
where again $\tilde{\chi}( \cdot )$ is a universal function. This
relation is perfectly satisfied, cf. Figure \ref{fig2a}.

\subsection*{Critical exponents $\tau$ and $\sigma$}
\begin{figure}
\includegraphics*[width=0.47\textwidth]{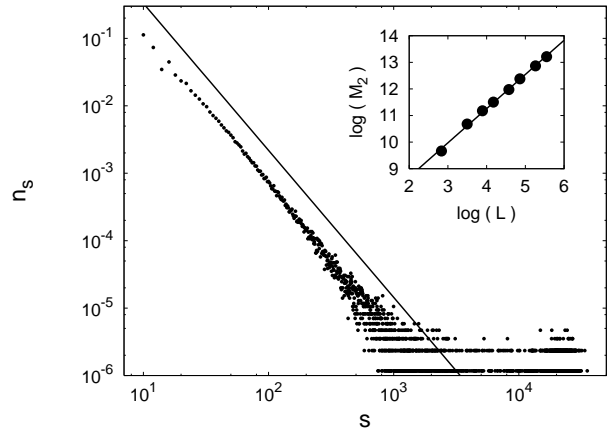}
\caption{The main graph shows the distribution of the cluster size
$n_s(p_c)$ at the critical point $p_c$ and for $L=257$. The
distribution is extracted from $10^3$ frozen configurations. As
expected this distribution follows a power law $n_s \sim
s^{-\tau}$, the solid line plotted here corresponds to
$\tau=2.19$. The inset shows the second moment of the
distribution of the cluster size $M_2$ at the critical point as a
function of the linear size of the lattice. $M_2$ increases as a
power of the linear size $L$ of the lattice with exponent
$\gamma/\nu = 1.28(2)$ (solid line).} \label{fig4}
\end{figure}

\begin{figure}
\includegraphics*[width=0.47\textwidth]{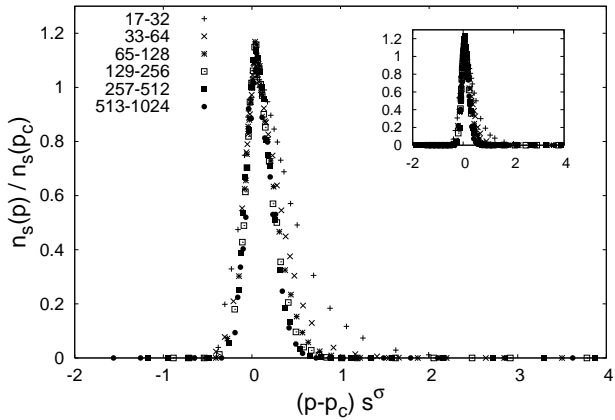}
\caption{Numerical test of the validity of equation
(\ref{eq:dist}). We plot the ratio $n_s(p)/n_s(p_c)$, where $n_s$
is the distribution of the cluster size, as a function of the
rescaled quantity $(p-p_c)s^{\sigma}$ with $\sigma$ chosen as $0.41$
from (\ref{eq11}) knowing $\beta$ and $\gamma$. The main plot refers to
$L=257$, the inset to $L=129$.} \label{fig4a}
\end{figure}

Furthermore, we consider the probability distribution  of having
$n_s$ clusters with $s$ negative links. $n_s$ is given by the
ratio of the number of clusters of size $s$ to the total number of
clusters. As it is known from percolation theory, $n_s$ should
satisfy
\begin{equation}
n_s = s^{-\tau} \tilde{n} \left[ \left(p-p_c \right) s^{\sigma} \right]  \;\;\; ,
\label{eq:dist}
\end{equation}
where $\tau$ and $\sigma$ are
critical exponents and $\tilde{n}( \cdot )$ is a universal
function. We can determine the critical exponent $\tau$ (also
called Fisher exponent) by plotting the distribution of the
cluster sizes $s$ at the critical point $p_c$. This distribution
is calculated in  Figure \ref{fig4} for $L=257$. The distribution
fits with a power law $s^{-\tau}$, here plotted as solid line with
$\tau=2.19$. We checked that the same exponent fits also for
smaller values of $L$.
\\
In Figure \ref{fig4a} we numerically determine the critical exponent
$\sigma$. The figure shows the plot of the ratio
$n_s(p)/n_s(p_c)$. Using the rescaled variable $(p-p_c)
s^{\sigma}$ as abscissa, all the curves corresponding to different
values of the cluster size $s$ ( $2^4 < s \leq 2^5$, $2^5 < s \leq
2^6$, $2^6 < s \leq 2^7$, etc\ldots ) collapse to a single
function for $\sigma=0.41$, as expected, see e.g. \cite{staufer}.

\subsection*{Check of the critical exponent $\gamma$}
{}From the cluster distribution it is possible to verify the
value of the critical exponent $\gamma$ that was calculated
before. It is known that the second moment of the distribution of
the cluster size
\begin{equation}
M_2= \sum_s s^2 n_s
\label{eq:second_moment}
\end{equation}
should scale according to $M_2 \sim L^{\gamma/\nu}$ at the
critical point $p_c$. The numerical value found for the ratio
$\gamma/\nu=1.28(2)$ is consistent with the former one obtained
via the susceptibility (see the inset of Figure \ref{fig4}).

\subsection*{Fractal dimension $d_f$}
\begin{figure}
\includegraphics*[width=0.47\textwidth]{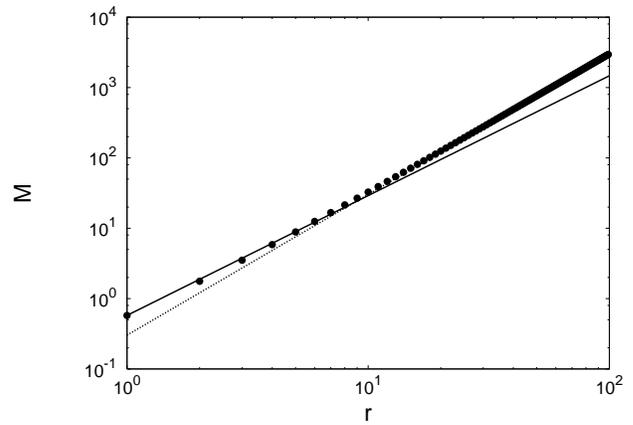}
\caption{Mass of negative links $M$ inside a triangular box with
$r$ links per side. The mass is measured on a triangular lattice
with $L=257$ and periodic boundary conditions. Each point is given
by the average over $10^3$ frozen configurations, while the number
of boxes considered for each $r$ is $10^2$.  The numerical results
fit a power law with $d_f=1.703$ (solid line) and $d=2$ (dotted
line).} \label{fig5}
\end{figure}

The percolating cluster can be further characterized in terms
of a fractal dimension $d_f$. The fractal dimension $d_f$ is
easily computed using a box counting method \cite{havlin}. We
count how many negative links $M(r)$ belong to a triangular box
with $r$ links per side. The measurement is performed for $L=257$
at the critical point $p_c$. We analyze $10^3$ frozen
configurations, for each of them we consider $10^2$ different
boxes. The mass $M(r)$ per box is plotted in Figure \ref{fig5}. As
expected we observe the crossover phenomenon
\begin{equation}
M(r) =
\left\{
\begin{array}{ll}
r^{d_f} & \textrm{ , if }r \ll \xi\\
r^{d} & \textrm{ , if }r \gg \xi
\end{array}
\right.
\;\;\; ,
\label{eq:fracd2}
\end{equation}
where $d=2$ in this case. The numerical results fit with a power
law with $d_f=1.703$ and $d=2$, respectively.
\subsection*{Time to reach the frozen states}
\begin{figure}
\includegraphics*[width=0.47\textwidth]{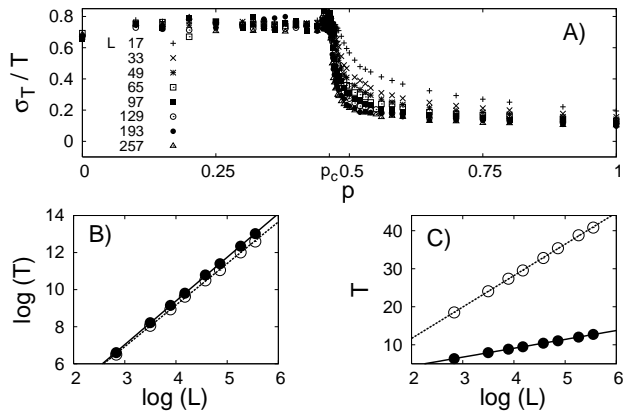}
\caption{Average time $T$ needed for reaching a frozen
configuration. The datasets are the same as in Figure \ref{fig1}.
In A) we plot the standard variance $\sigma_T$ over the average
value $T$ as a function of $p$ and for different values of $L$.
The peaks of this ratio are all around the critical point $p_c$.
In B) we show a log-log plot of  $T$ as a function of the linear
size of the lattice $L$ at the critical point $p_c$
(full circles) and at $p=1/3$ (open circles). It implies that $T
\sim L^{z}$ for $p \leq p_c$. We have $z=2.36(1)$ (solid line) at
$p_c$, while $z=2.24(2)$ at $p=1/3$ . In C) the plot of $T$
versus the logarithm of $L$ for $p=1$ (full circles) and
for $p=3/4$ (open circles) shows that $T \sim \log{\left(L\right)}$
for $p>p_c$. The numerical simulations are the same as
in Figure \ref{fig:cum}.} \label{fig6}
\end{figure}

So far we characterized the geometrical properties of the final
absorbing configuration. As a next step we focus on the dynamical
features and determine the time the system needs to reach the
final frozen configurations. In Figure \ref{fig6}A) we plot the
ratio of the variance $\sigma_T$ and its average value $T$ as a
function of $p$. It is interesting to notice that $\sigma_T/T$ has
a maximum around $p_c$. Below the critical point $p_c$, the
relaxation time $T$ is governed by a power-law relation in the
linear size of the lattice $L$: $T \sim L^z$. Our numerical
analysis reveals $z=2.36(1)$ at $p_c$ and $z=2.24(1)$ at $p=1/3$
[see Figure \ref{fig6}B)]. Furthermore, above the critical point
$p_c$,  $T$ grows logarithmically with $L$: $T \sim \log{(L)}$
[see Figure \ref{fig6}C), where we show the $L$ dependence of $T$
for $p=3/4$ and for $p=1$]. The different size dependence of the
time the system needs to reach a frozen configuration in both
phases can be understood in qualitative terms. For  $p
> p_c$ there is a lower probability for having negative links. The
stable configurations forming out of these links are mainly local
objects like star-like configurations which do not percolate
through the lattice [see Figure \ref{fig:expl1}C)]. Therefore the
time to reach such a state characterized by local objects depends
weakly on the system size, that is logarithmically. For $p\leq p_c$ there is a high probability of having negative
links. However, here large loops that finally percolate through
the lattice make up the stable configurations [see Figure
\ref{fig:expl1}A) and B)]. So the time to reach this configuration
is more sensitive to the system size, it grows like a power of the
size.

\subsection*{Universality class of triad dynamics in two dimensions}
We list the critical exponents for the phase transition between
different absorbing states in Table \ref{table}.
\begin{table}
\begin{center}
\begin{tabular}{|l|r|}
\hline
$p_c$ & $0.4625(5)$\\
\hline
$\beta$ & $0.46(5)$\\
\hline
$\gamma$ & $2.0(3)$\\
\hline
$\nu$ & $1.6(2)$\\
\hline
$\sigma$ & $0.41(6)$\\
\hline
$\tau$ & $2.19(1)$\\
\hline
$d_f$ & $1.703(3)$\\
\hline
$z$ & $2.36(1)$\\
\hline
\end{tabular}
\end{center}
\caption{Critical threshold and critical exponents numerically
determined for the absorbing configurations of triad dynamics on
two-dimensional triangular lattices.} \label{table}
\end{table}
The critical exponents satisfy the known hyperscaling relations.
For example, it is known from percolation theory that the critical
exponents $\beta$, $\gamma$, $\tau$ and $\sigma$ are related by
\begin{equation}
\tau=\frac{1}{2} \left[ 5- \frac{\gamma-\beta}{\gamma+\beta} \right]
\label{eq:tau}
\end{equation}
and
\begin{equation}
\sigma=\frac{1}{\gamma+\beta}\;\;\; .
\label{eq:sigma}
\end{equation}\label{eq11}
Calculating $\tau$ and $\sigma$ from the former equations, using
the numerically obtained values of $\beta$ and $\gamma$, we find
$\tau=2.19(1)$ and $\sigma=0.41(6)$. Both values are consistent
with Figures \ref{fig4} and  \ref{fig4a}.
\\
The fractal dimension is related to the ratio $\beta/\nu$ by the hyperscaling relation
\begin{equation}
d_f = d - \frac{\beta}{\nu}\;.
\label{eq:fracd1}
\end{equation}
{}From this relation we find $d_f=1.703(3)$ in agreement with
Figure \ref{fig5}.
\\
The errors of $p_c$ and of the critical exponents as indicated in
Table \ref{table} arise as follows. The critical point $p_c$ is
directly estimated from Figure \ref{fig:cum}, verified in Figure
\ref{fig1} and supported by Figure \ref{fig5}. The only source of
error here is given by the step size used for varying the
propensity parameter $p$. Of course this step size is related to
the parameters of the simulations. It can be decreased by
increasing the system size and the total number of simulations.
\\
The critical exponent $\beta/\nu$ of Figure \ref{fig1}B) , $1/\nu$
of Figure \ref{fig:cum}, $\gamma/\nu$ of Figures \ref{fig2a} and
\ref{fig4}, and $z$ of Figure \ref{fig6} have errors due to the
linear fit in a double-logarithmic plane. In principle one should
also account for the propagation of the error entering the value
of the critical point $p_c=0.462(5)$, which we here have neglected.
\\
In order to check the validity of the hyperscaling relations we
evaluate the corrections to all derived quantities ($\tau, \sigma$ and  $d_f$) by using the
standard formula for error propagation. The hyperscaling relations
are then said to be satisfied if they hold within these error
bars.
\\
To our knowledge the critical exponents calculated so far are new
\cite{staufer,havlin,odor}. Therefore, the percolation transition between
different absorbing states of triad dynamics on two-dimensional
lattices can be described by standard percolation theory, but the
transition seems to belong to a new universality class.

\section{Tetrad dynamics}\label{sec4}

\begin{figure}
\Large{A)}
\\
\includegraphics*[width=0.39\textwidth]{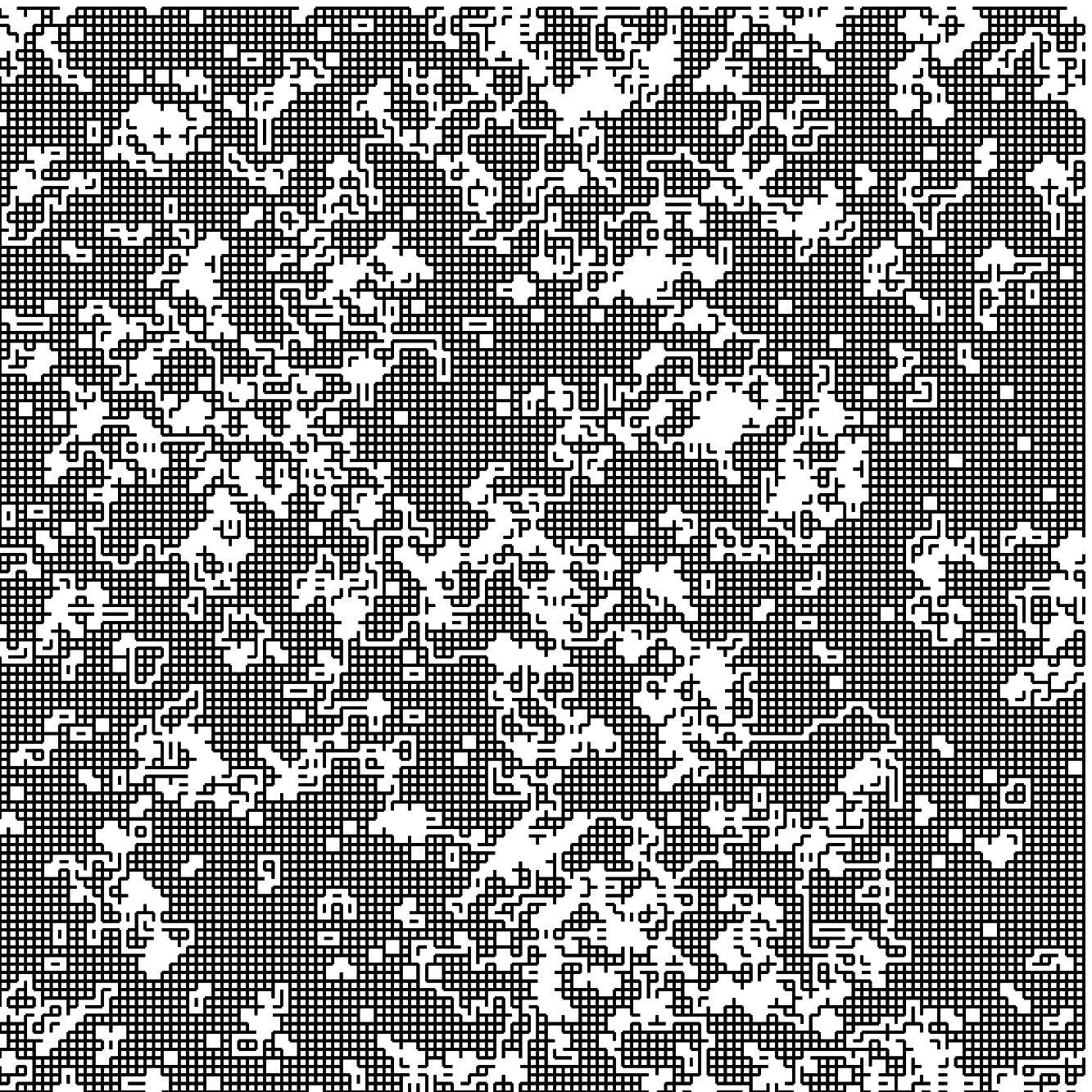}\\
\vspace{0.3cm}
\Large{B)}
\\
\includegraphics*[width=0.39\textwidth]{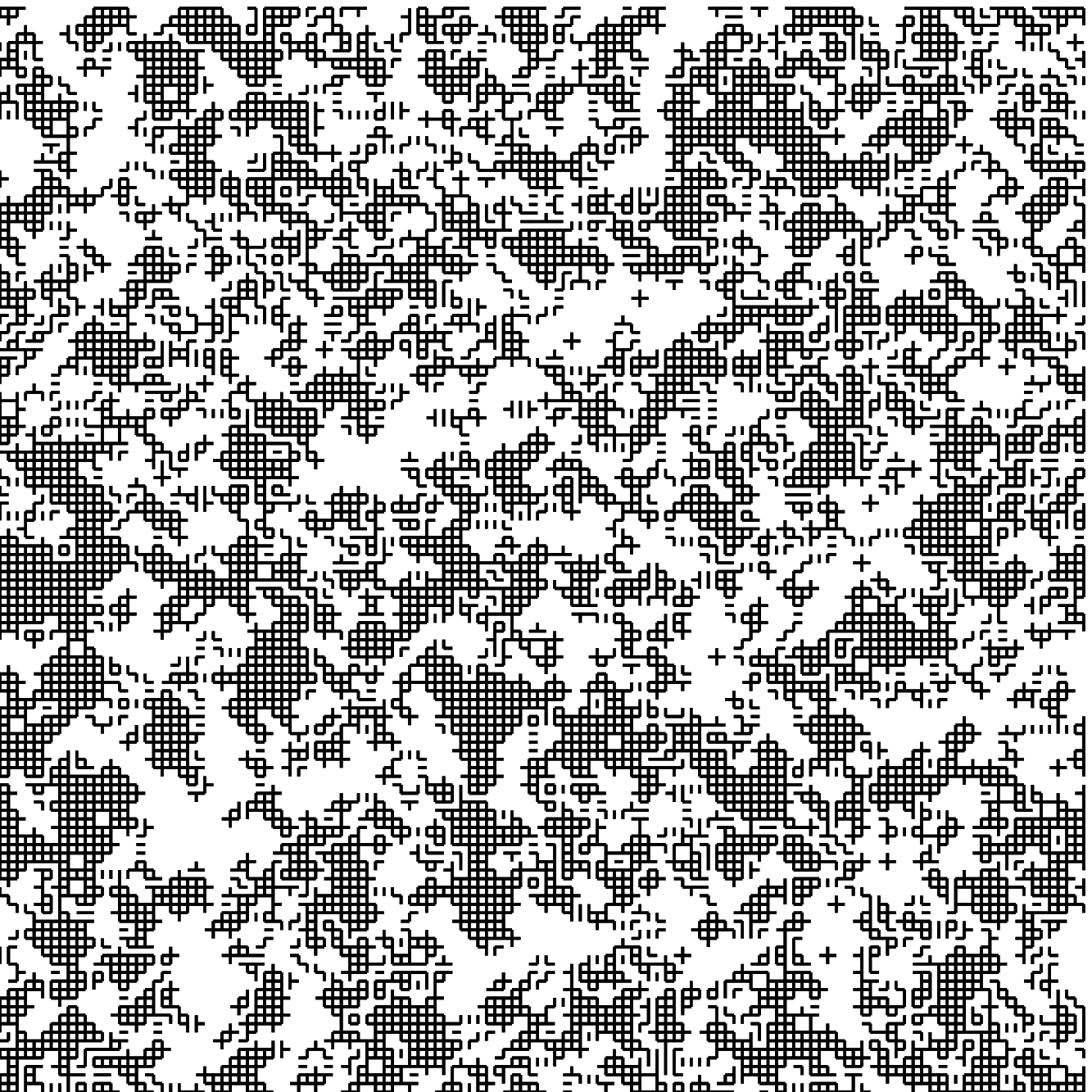}\\
\vspace{0.3cm}
\Large{C)}
\\
\includegraphics*[width=0.39\textwidth]{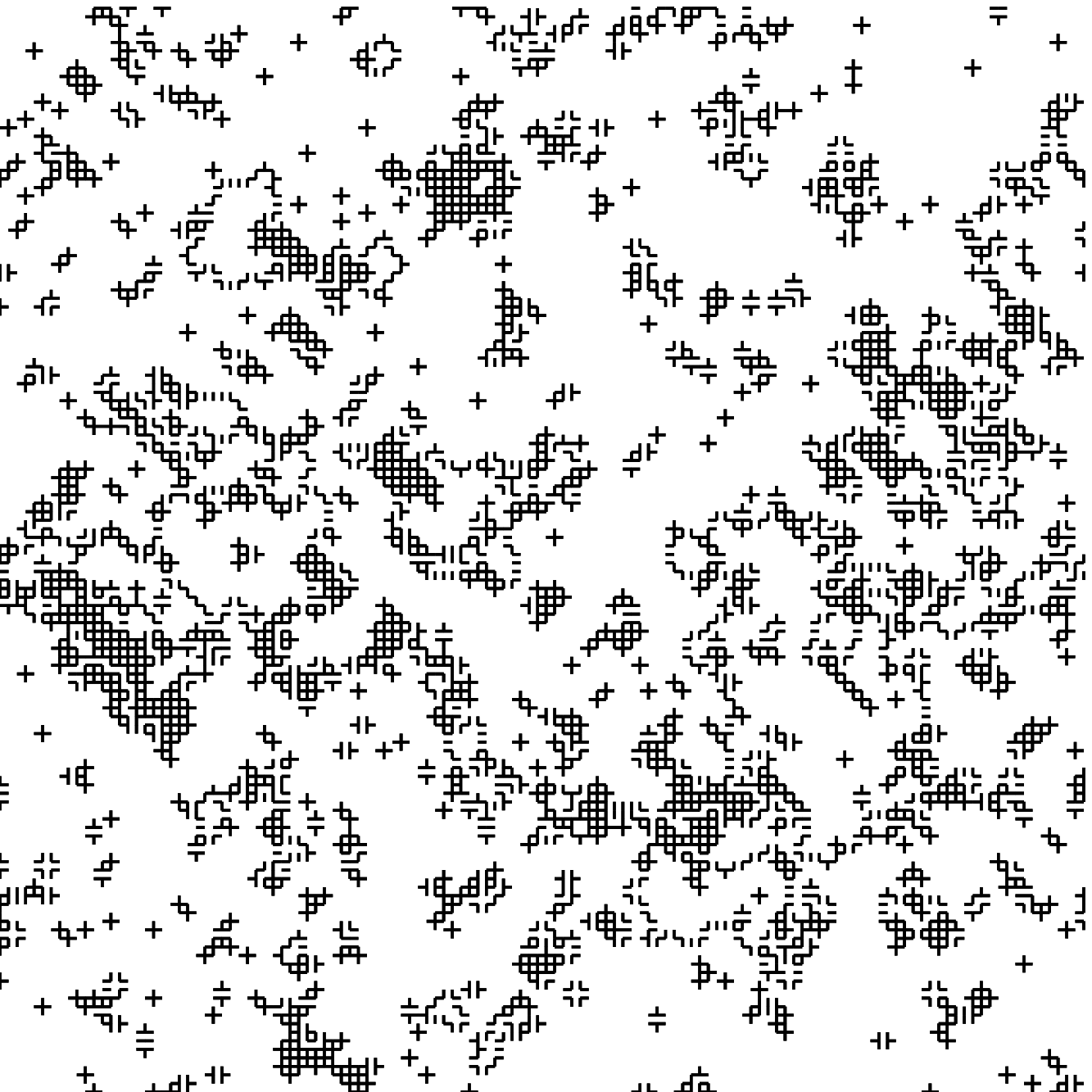}
\caption{Typical frozen configurations for a square lattice
with periodic boundary conditions, with linear size $L=128$ and
for different values of $p$: A) $p=0.48$ , B) $p=p_c=0.5$ , C) $p=0.52$.
In these plot are shown only the negative links.}
\label{4fig:expl}
\end{figure}

\begin{figure}
\includegraphics*[width=0.47\textwidth]{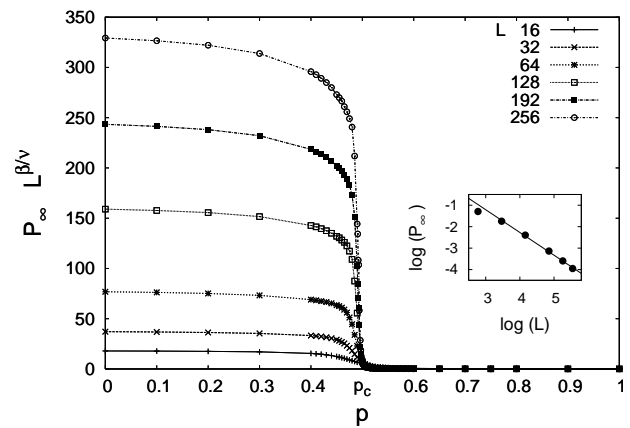}
\caption{Probability $P_\infty$ that a link belongs to the largest
cluster as a function of $p$ and for different linear sizes of the
lattice. The ratio of critical exponents  $\beta/\nu = 1.05(2)$ at
the critical point $p_c=0.5$ as is shown in the inset.}
\label{4fig1}
\end{figure}

\begin{figure}
\includegraphics*[width=0.47\textwidth]{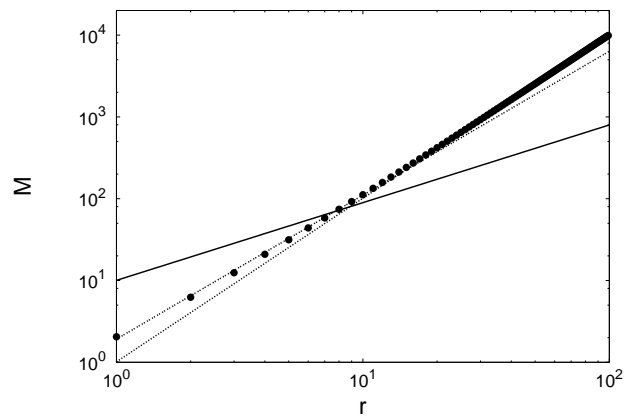}
\caption{Mass of negative links $M$ inside a quadratic box with
$r$ links per side. This measure is used on a square lattice with
$L=256$ and periodic boundary conditions. Each point is given by
the average over $10^3$ frozen configurations, while the number of
boxes considered at each $r$ is $10^3$. The solid line has the
slope $0.95$, a value that is expected by the former measure of
$\beta/\nu$. Actually it is not consistent with the direct fit of
the data from which  we find $d_f=1.76(2)$ (dashed line). The
dotted line has slope $d=2$.} \label{4fig2}
\end{figure}

The notion of social balance can be extended to any geometric
figure with $k$ links (in graph theory denoted as $k$-cycles)
\cite{radicchi,cartwright}. As it turned out in \cite{radicchi}
for an all-to-all topology and a diluted topology, generalizing
the triad dynamics to a $k$-cycle dynamics leads to qualitative
differences in the phase structure of which the main differences
are due to $k$ being even rather than being larger than three. In
this section we therefore focus on the case of $k=4$ and call the
$4$-cycles tetrads defined on square lattices.  According to the
number of negative links belonging to a particular tetrad, we
distinguish five types of tetrads: $\square_0$ , $\square_1$ ,
$\square_2$ , $\square_3$ , $\square_4$ . The updating rules of
the unconstrained dynamics for triads of (\ref{eq:triad_rules})
are naturally extended to tetrads in the following way:
\begin{equation}
\square_0 \;\;  \xleftarrow[p]{} \square_1  \xrightarrow[1-p]{} \;\; \square_2 \;\;  \xleftarrow[p]{}
\square_3  \xrightarrow[1-p]{} \;\; \square_4 \;\;\; .
\label{eq:tetriad_rules}
\end{equation}
The local tetrad dynamics is then applied to square lattices with
periodic boundary conditions. $L$ denotes the linear size of the
lattice, i.e. the number of sites per row or per column. The total
number of sites of the lattice is $N=L^2$, the total number of
links $M=2N$, and the total number of tetrads $N_{\square}=N$.
\\
\vspace{0.5cm}
\\
Similarly to the case of triad dynamics on two-dimensional
triangular lattices, at a first glance there seems to be a
percolation transition also in case of tetrad dynamics. The
critical point $p_c$ should be equal to $p_c=0.5$ due to the
symmetry of the system (\ref{eq:tetriad_rules}) under the
simultaneous transformation $\sigma_i \to -\sigma_i \;\; \forall
i$ and $p \to 1-p$ (see Figure \ref{4fig:expl}).
\\
However, differently from the triad dynamics, no scaling relations
hold in the vicinity of the transition at which again an infinite
cluster of negative links forms. The hyperscaling relations are
violated. For example, the relation (\ref{eq:fracd1}) does not
hold because one can measure $\beta/\nu=1.05(2)$ (Figure
\ref{4fig1}) and expect to have $d_f=0.95(2)$, while it is
directly seen from Figure \ref{4fig2} that $d_f=1.76(2)$.

\vspace{0.5cm}

\begin{figure}
\includegraphics*[width=0.47\textwidth]{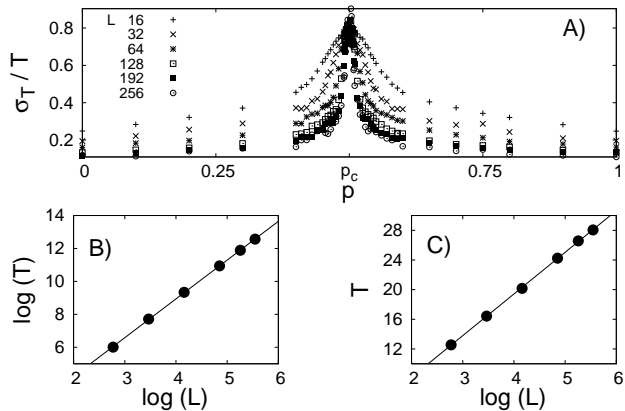}
\caption{Average time $T$ needed for reaching a frozen
configuration. The simulations are the same as in Figure \ref{4fig1}.
In A) we plot the standard variance $\sigma_T$ over the average
value $T$ as a function of $p$ and for different values of $L$.
The peaks of this ratio are all in the vicinity of the critical
point $p_c$. In B) we show a log-log plot of  $T$ as a function of
the linear size of the lattice $L$ at the critical point leading
to $T \sim L^{z}$ with $z=2.35(1)$. In C) the plot of $T$ versus
the logarithm of $L$ for $p=1$ shows that
$T\sim\log{\left(L\right)}$.} \label{4fig3}
\end{figure}

Tetrad dynamics resembles triad dynamics. Each unstable tetrad
either diffuses, or two unstable tetrads annihilate if they meet.
Finite-size systems always reach a frozen configuration within a
finite time, so we can measure this time $T$. As we can see from
Figure \ref{4fig3}A), we find a symmetric behavior around the
critical point $p_c$. At the critical point $T$ scales according
to $T \sim L^z$ with $z=2.35(1)$ [see Figure \ref{4fig3}B)], this
value is actually consistent with the one found for the triad
dynamics. Away from the critical point $T \sim \log{(L)}$ [see
Figure \ref{4fig3}C)].

\section{Summary and Conclusions}\label{sec5}
The driving force in triad dynamics is the reduction of the number
of frustrated triads. A state of zero frustration is called a
state of social balance. Imposed on a triangular lattice,
frustrated triads can diffuse or annihilate each other. Depending
on the value of the propensity parameter the final absorbing state
can be characterized by the absence ($p>p_c$) or presence ($p\leq
p_c$) of an infinite cluster of negative links. The time to reach
the frozen configurations scales with the system size in a way
that further characterizes the phases: it scales logarithmically
for $p > p_c$ and power-like for $p\leq p_c$. The critical
exponents $\nu$, $\beta$, $\gamma$, $\tau$, and $\sigma$ as well
as the fractal dimension $d_f$ satisfy hyperscaling relations
within the error bars. The values of these exponents seem to
characterize a new universality class. The essential difference
that we observe for tetrad dynamics on a square lattice is the
symmetry between the absorbing states in the different phases.
Again, an infinite cluster of negative links forms for $p\leq
p_c$, but the time to reach the frozen configurations shows the
same dependence on the system size in both phases, with the only
exception at the transition point. The percolation picture breaks
down in the sense that no scaling and therefore no hyperscaling
relations are satisfied.

\end{document}